\documentclass{article}
\usepackage{ijcai19}

\usepackage{times}
\usepackage{soul}
\usepackage{url}
\usepackage[hidelinks]{hyperref}
\usepackage[utf8]{inputenc}
\usepackage[small]{caption}
\usepackage{graphicx}
\usepackage{amsmath}
\usepackage{amssymb}
\usepackage{amsthm}
\usepackage{booktabs}
\usepackage{algorithm}
\usepackage{algorithmic}
\usepackage{amsthm}
\usepackage{xcolor}
\usepackage{verbatim}
\newcommand{\eg}{\emph{e.g. }}

\newcommand{\ie}{\emph{i.e. }}
\newcommand{\etc}{\emph{etc}}

\title{Automatically Searching for U-Net Image Translator Architecture}

\author{
Han Shu
\and
Yunhe Wang
\affiliations
Huawei Noah's Ark Lab
\emails
\{han.shu, yunhe.wang\}@huawei.com
}

\begin{document}

\maketitle

\begin{abstract}
 Image translators have been successfully applied to many important low level image processing tasks. However, classical network architecture of image translator like U-Net, is borrowed from other vision tasks like biomedical image segmentation. This straightforward adaptation may not be optimal and could cause redundancy in the network structure. In this paper, we propose an automatic architecture searching method for image translator. By utilizing evolutionary algorithm, we investigate a more efficient network architecture which costs less computation resources and achieves better performance than the original one. Extensive qualitative and quantitative experiments are conducted to demonstrate the effectiveness of the proposed method. Moreover, we transplant the searched network architecture to other datasets which are not involved in the architecture searching procedure. Efficiency of the searched architecture on these datasets further demonstrates the generalization of the method. 
\end{abstract}

\section{Introduction}

Image-to-image translation has been a fundamental research field of computer vision. Image-to-image translation tasks such as domain translation~\cite{cyclegan}, image super-resolution~\cite{SRGAN}, image colorization\cite{colorpix} and image denoising~\cite{chen2018image} are widely involved in image processing of common applications. To translate image from one domain to another,~\cite{larsson2016learning}  predicts per-pixel color histogram as an intermediate output to help the translation procedure. \cite{zhang2016colorful} posts it as a classification task and uses class-rebalancing at training time to increase the reality of the results. The pix2pix method\cite{pix2pix},  firstly leverages generative adversarial networks in a conditional setting to solve the image-to-image translation issue. ~\cite{chen2017photographic} suggest that adversarial training might be unstable and prone to failure for high-resolution image generation tasks. Instead, they adopt a modified perceptual loss\cite{johnson2016perceptual} to synthesize images, which are high resolution but often lack fine details and realistic textures. ~\cite{wang2018high} presented a novel adversarial loss as well as a coarse-to-fine generator and a multi-scale discriminator to address the high-resolution image-to-image translation problem. These previous methods have made great progress and achieves good performance on image-to-image translation tasks. However, the network architectures become more complicated and consume more computation cost.

In addition, most of them studied the loss functions, weight normalization and training techniques, which never deal with the architectures of translator explicitly. The design of deep neural networks can have enormous influence on its performance. ResNet~\cite{he2016deep} achieved much higher accuracy than VGGNet~\cite{simonyan2014very} on image classification but required fewer parameters by introducing the residual blocks. ~\cite{dcgan} introduced deep convolutional generative adversarial networks which can generate better images. Although manually designed network architectures can achieve good performance in various tasks, recent progress have proved that architectures that built by automatically searching can outperform hand-craft structures. ~\cite{zoph2016neural} exploited reinforcement learning to generate architectures which can achieve better performance than the human-invented networks with less parameters. ~\cite{real2017large} employed evolutionary algorithms to search the architecture of neural network. ~\cite{liu2018progressive} proposed a more efficient method for learning the structure of CNNs utilizing a sequential model-based optimization strategy. ~\cite{liu2018darts} introduced a continuous relaxation to represent the architecture to allow efficient searching using gradient descent. 
\begin{figure*}[t]
	\centering
	\includegraphics[width=0.9\linewidth]{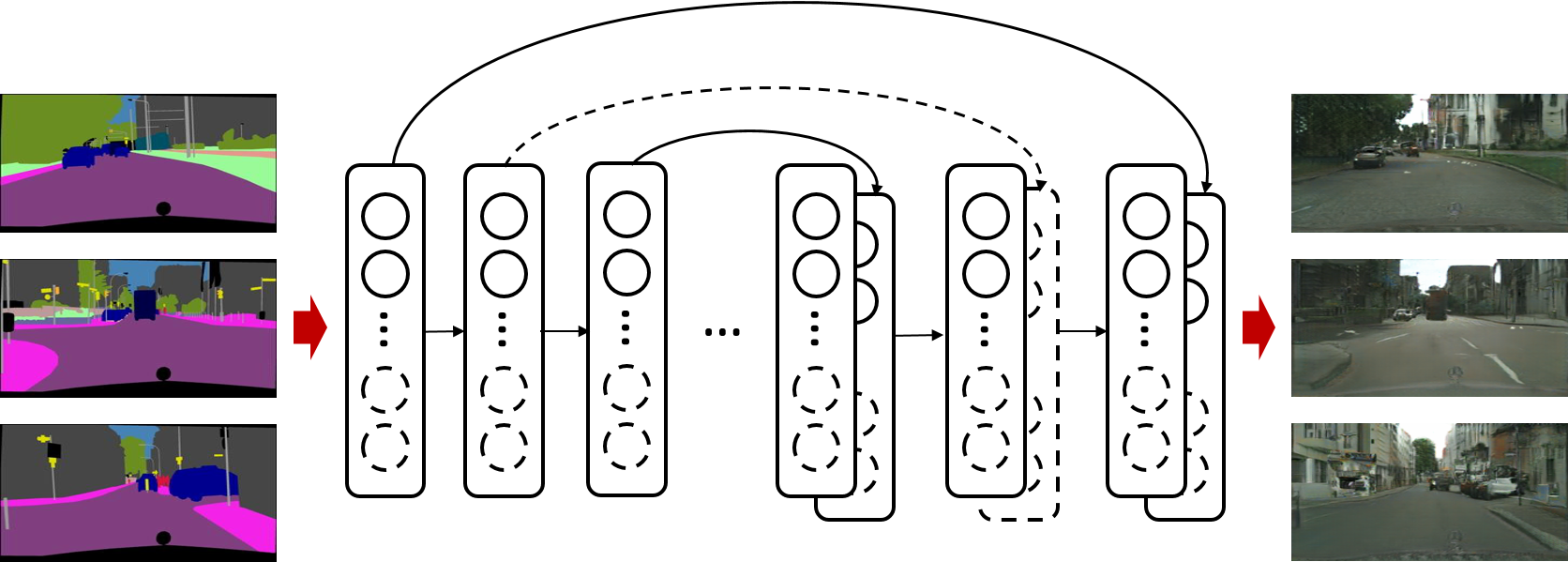}
	\caption{The framework of the proposed method for automatically searching the architecture of the image-to-image translation task. We search the channel numbers of each convolutional and deconvolutional layer, and the connection between the mirrored layers. The solid line represents the exsitance of the skip connection while the dotted line represents the opposite.}
	\label{Fig:framework}
\end{figure*}

There are major differences between the visual recognition task and image translation task. First, generative model produced by an adversarial training process has to further deal with a discriminator model. While the traditional visual recognition only requests a discriminator. Second, the architecture of an image generator is different from that of an image classifier.   Some generators take U-Net~\cite{unet} as the basic architecture, which has a symmetrical structure and skip connections between the convolutional layers and deconvolutional layers. The searching space of U-net is fundamentally different from that of ordinary deep neural networks for classification. To this end, it is necessary to tailor a new neural architecture search method for generator in the image translation task.

In this paper, we propose a novel framework to automatically search the architecture of  image translator utilizing the evolutionary algorithm. We use the U-net as the backbone and explore the channel numbers of each convolutional layer and deconvolutional layers as well as the essentiality of every skip connection in U-Net.  We carefully design the genetic algorithm to apply it to the architecture search of U-Net. The network search is conducted on cityscapes dataset. Then, the searched architecture is transplanted to other benchmark dataset of image-to-image translation. Extensive experiments demonstrate the effectiveness of the proposed method.
\section{Preliminaries}
Here we first briefly introduce the image-to-image translation task with generative adversarial networks and the genetic algorithm for automatically searching the network architecture.
\subsection{Image-to-Image Translation}
image-to-image translation can be taken as a per-pixel classification or regression problem \cite{colorpix}. Recently, \cite{pix2pix} used generative adversarial networks to solve this problem. 
In an image-to-image translation task, such as semantic map to street view on cityscapes dataset, the objective function within the framework of GANs can be written as
\begin{equation}
\begin{aligned}
\mathcal{L}_{GAN}(G,D) =  \mathbb{E}_{y\sim p_{data}(y)} \left[\log D(y)\right] &\\
+\mathbb{E}_{x\sim p_{data}(x)}\left[\log(1-D(G(x)))\right],&
\end{aligned}
\label{Fcn:GAN}
\end{equation} 
where $X$ represents the source domain (\eg semantic map), and $Y$ represents the target domain (\eg street view). Denote the training sample in domain $X$ as ${x_1,x_2,...,x_n}$, and the corresponding images in domain $Y$ as ${y_1,y_2,...,y_n}$. The generator $G$ transforms images from source domain $X$ to target domain $Y$(\ie $G:x-y$), while the discriminator $D$ distinguishes which domain the input comes from. The task of domain transfer using generator and discriminator is formulated as a minimax problem. The generator $G$ is optimized to generated images to fool the discriminator $D$. Whereas the discriminator $D$ is optimized to distinguish which domain the image come from.

In the task of paired image-to-image. translation \cite{pix2pix}, except for the regular loss for generator, and the conventional $\ell_1$ loss between fake images from the generator and real target images is often adopted, \ie   
\begin{equation}
\mathcal{L}_{L_1}(G) = \mathbb{E}_{y\sim p_{data}(y)} \left[ {\|y-G(x)\|}_{1}\right].
\label{Fcn:l1loss}
\end{equation}
Thus, the objective function for an image-to-image translator~\cite{pix2pix} can be formulated as
\begin{equation}
\min_G \max_D \mathcal {L}_{GAN}(G,D) + \lambda \mathcal{L}_{L_1}(G).
\label{Fcn:total generator}
\end{equation}
With the help of minimax design of GAN and $\ell_1$ loss, the generator can achieve satisfactory performance on image-to-image. translation task. However, the generator architecture U-net~\cite{unet} is directly borrowed from biomedical image segmentation  and may not be the optimal choice for this supervised task.

\subsection{GA for Network Architecture Search}
Neural architecture search has attracted much attention in the field of deep learning. There are two popular ways to do the search, evolutional algorithm and reinforcement learning. In this paper, we adopt evolutional algorithm to explore the architecture space as suggested in~\cite{real2018regularized}.

In Genetic Algorithm (GA) based network architecture search, we need to maintain a population of individuals, each of which represents a certain network architecture, to realize the searching process. The population in the current generation are regarded as parents, who breed next generation through three kinds of operations including selection, crossover and mutation, with the expectation that the subsequent offspring perform better than the parents. After enough number of generations, remaining offspring would suit better for the designed task. Each individual $b_{j}$ is assigned with a probability  $Pr(b_{j})$ through a roulette algorithm according to its fitness:  
\begin{equation}
\mathcal Pr(b_{j}) = f(b_{j})/\sum_{k=1}^{K}f(b_{k})
\label{Fcn:prob}
\end{equation}
where $K$ is the number of individuals in the population, $b_{j}$ is the $j$-th individual for encoding a specific neural architecture, and $f(b_{j})$ is the fitness of $b_{j}$. Then three operations will be executed  with the probability of $s_1$, $s_2$ and $s_3$ respectively, and $s_1+s_2+s_3=1$.
 
\textbf{Selection:} Given probability $s_1$, selection is conducted. An individual selected according to Fcn.\ref{Fcn:prob} is directly copied as an offspring. Individual with higher fitness has more chance to be preserved. 
 
\textbf{Crossover:} Given probability $s_2$, crossover is conducted. Two individuals from parent generation are selected according to Fcn.\ref{Fcn:prob}. Random piece of parents will be swapped to generate two new offsprings. This operation is to integrate excellent genetic fragments of the parents. 
 
\textbf{Mutation:} Given probability $s_3$, mutation is conducted. One individual from parent generation is selected according to Fcn.\ref{Fcn:prob}. Mutation randomly changes a random piece in the parent individual to produce an offspring. The common mutation operation for binary encoding is XOR operation. This operation increase the diversity of the population.  

By iteratively employing these three genetic operations, initial population will be updated efficiently until the maximum iteration number is achieved. After obtaining the individual with optimal fitness, we can get a new architecture of network. The key points for applying the genetic algorithm to network architecture search is to design the representation code for each individual and the fitness which evaluates the performance of each individual of a specific task.

\section{Method}
In this part, we will introduce the details of the proposed method including the search space, the representation of the individual architecture and how evolutional algorithm is applied during the search process. We use U-Net as the backbone to illustrate this part.

\subsection{Search Space of Architecture}
U-Net \cite{unet} was first proposed for biomedical image segmentation while widely used in image-to-image translation task. U-Net is similar to an encoder-decoder network composed of sequential convolutional layers and deconvolutional layers. Through the network, feature maps are empirically down sampled with more channels, and are up sampled to the original size in the reverse manner~\cite{pix2pix}. Different from the conventional encoder-decoder structure, U-Net has skip connections between mirrored layers in the encoder and decoder stacks. In this work, we follow the encoder-decoder fashion, but explore channel numbers of each convolutional and deconvolutional layers as well as the essentiality of each skip connection. In addition, the sizes of filters in all convolutional layers are $4\times4$ with a stride of $2$ for extracting visual features with considerable  receptive fields and ensuring the consistency of the encoder-decoder architecture~\cite{unet}, which will not to be searched. Therefore, the search space of these configurations is defined as follows
\begin{equation}
\centering
S_{1}= \{64,128,256,512\}, \quad S_{2}= \{1, 0\},
\end{equation}
where $S_{1}$ the set of available choices of channel numbers, which includes all the optional output channel numbers of each convolutional layers. Due to the symmetry of U-Net, we only have to operate on the first half of the network to determine the whole one. $S_{2}$ is the set of available choices of skip connection. $1$ is to keep a certain skip connection while $0$ is to remove the skip connection. For a U-Net generator used in \cite{pix2pix}, there are $8$ pairs of convolutional and deconvlutional layers with $7$ skip connections. The size of search space for this architecture is $4^8\times2^7=8,388,608$, which cannot be efficiently optimized by conventional methods.

\subsection{Representation of Search Architecture}
\begin{figure}[t]
\centering
\includegraphics[width=1\linewidth]{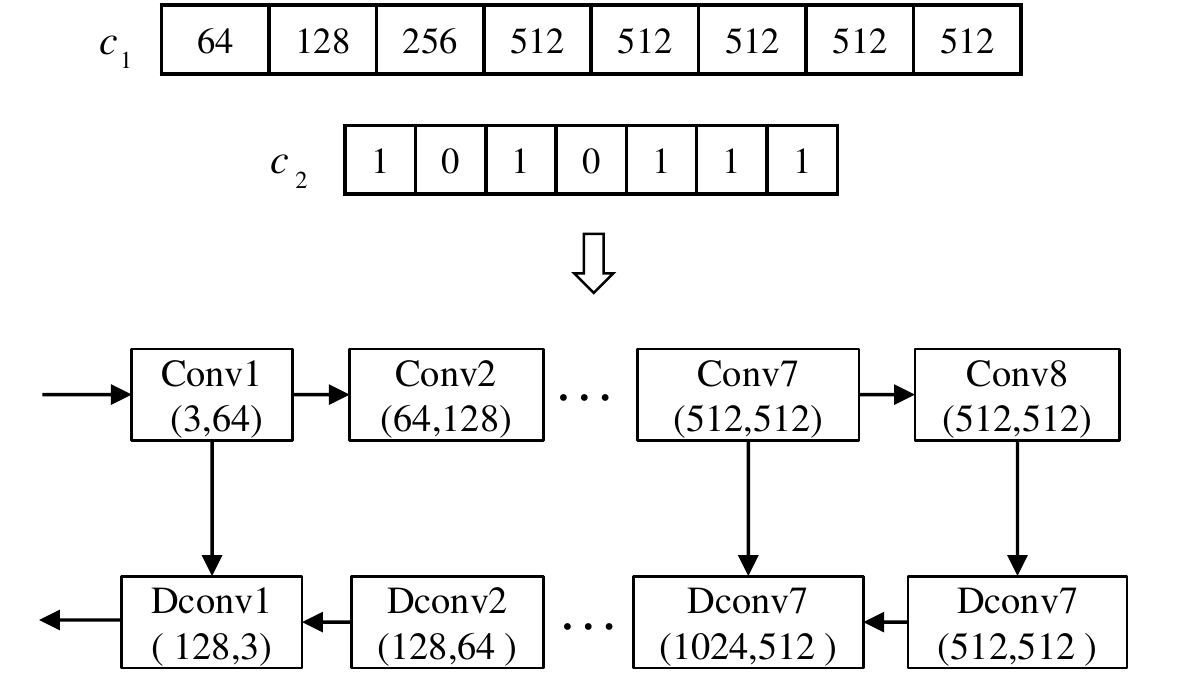} 
\caption{The decoding procedure of the proposed method for reconstructing a U-Net using an individual. The channel number of each layer and the connections will be recognized from the given individual. Then, the number of convolution filters in each layer can be set accordingly.}
\label{Fig:construct}
\vspace{-0.5em}
\end{figure}

We use two fixed-length codes to represent each variant of U-Net. $c_{1}$ represent the code for output channel numbers, which defines output channel numbers of the first half of convolutional layers. $c_{2}$ represent the code for skip connection, which determine whether to keep or remove each skip connection. $L_{c_{1}}$ is the length of $c_{1}$, and $L_{c_{2}}$ is the length of $c_{2}$. To produce initial individuals, a bootstrap sampling from $S_1$ is repeated for $L_{c_{1}}$ times to generate a single channel number code $c_{1}$ and a bootstrap sampling from $S_2$ is repeated for $L_{c_{2}}$ times to generate a single connection code $c_{2}$.   Every pair $(c_{1},c_{2})_{i}$ represents a particular individual in the search space. Taking original U-Net for an example,  $L_{c_{1}}=8$ and $L_{c_{1}}=7 $.  The decoding process of the proposed method for constructing a new network is shown in Fig.~\ref{Fig:construct}. First step is to determine the pair of code  $(c_{1},c_{2})_{i}$, either from initial individual or genetic operations of selection, crossover and mutation. Then, a new network with a specific architecture will be established according to the channel number code $c_{1}$. Finally, the numbers of input and output channels in each layer will be calculated based on the symmetric of U-Net and the connection code $c_{2}$ which indicates the remaining skip connections to formulate the resulting generator network.

\subsection{Evolutionary Strategy}
Usually, the generator $G$ and the discriminator $D$ in a conventional adversarial network are optimized alternatively, and the discriminator will be discarded after the training procedure. Therefore, we propose to utilize the genetic algorithm to update the population to get better architectures of the generator network, \ie $G$. The objective function for generator $G$ is formulated as
\begin{equation}
\begin{aligned}
\mathcal{L}_{GAN}(G) = \mathbb{E}_{x\sim p_{data}(x)}\left[\log(1-D(G(x)))\right].
\end{aligned}
\label{Fcn:GANG}
\end{equation}
To evaluate the performance of each individual architecture in each generation among the population, we have to consider both the model performance and computation efficiency. After some mini-epoch training of each individual, we conduct a validation procedure to evaluate each individual architecture. Individual fitness is calculated as
\begin{equation}
f(\mathbf{q}) = {\left( \mathcal{L}_{Gen}+\gamma p(\mathbf{q})\right)}^{-1},
\label{Fcn:fitness}
\end{equation}
where $p(\mathbf{q})$ is the FLOPs required by the given architecture for processing each input image, and 
\begin{equation}
\mathcal{L}_{Gen}= \mathcal{L}_{GAN}(G) + \lambda \mathcal{L}_{L_1}(G),
\label{Fcn:performloss}
\end{equation}
where $\mathcal{L}_{Gen}$ represents the cumulative  loss of the generator during validation procedure,  which reflects the image translation quality of the architecture. $\mathcal{L}_{Gen}$ is composed of a generator loss and the $\ell_1$ loss with a hyper parameter $\lambda$. The hyper parameter $\gamma$ is utilized for balancing the computation cost and model performance. A larger $\gamma$ represents that the evaluation of individuals focuses more on reducing the computation cost. Note that both FLOPs and generator loss are negative correlated with the fitness of network architecture, so a reciprocal operation is added in the fitness function.

\subsection{Search Procedure}
We adopt the classic genetic algorithm to conduct network architecture search described as Alg. \ref*{alg:genetic}. First, we initialize the first generation of population using random sampling from $S_1$ and $S_2$. For computation speed, we select subset of dataset for fast training and validation, and denote them as mini train set and mini validation set. Then, we apply every individual in the population for mini-train and mini-validation for calculating individual fitness utilizing Fcn .\ref{Fcn:fitness}. Then we can determine the probability for every individual to be selected. After that, we use three genetic operations, including selection, cross over and mutation to  generate offsprings in the iteration procedure. 

By iteratively updating a series of individuals in the population using the genetic operations for $T$ times, a new architecture of the generator network is discovered with a high probability to perform better than the original architecture in the image-to-image translation task. For an easier complementation, we adopt genetic algorithm separately for channel number code $c_1$ and connection code $c_2$.  For selection and cross over, regular strategy of genetic algorithm is adopted. As for mutation, we have to design special strategy to operate because the channel number code $c_1$ is not binary. Thus we cannot use XNOR operation directly for mutation. For mutation of channel number code $c_2$, we design a unique mutation operation: for an element corresponding to a channel number to be mutated, we remove the past channel number from set $S_1$ and randomly select one from the remaining numbers as the new channel number. 


\begin{algorithm}[t]
	\caption{Evolutionary search for the generator network.}
	\label{alg:genetic}
	\begin{algorithmic}[1]
		\REQUIRE Training set from two domains $X=\{x_i\}_{i=1}^n$ and $Y=\{y_i\}_{i=1}^n$ including $n$ paired images, the number of individuals $K$, the maximum iteration number $T$ in the genetic algorithm, $\lambda$, $\gamma$, and training parameters, \etc.
		\STATE Randomly initialize the population $P_0$ with $K$ individuals according to the search space;
		\FOR{$t = 1$ to $T$}
		\STATE Train each individual in $P_{t-1}$ on the mini-train set.
		\STATE Test each trained individual on the mini-validation set. 
		\STATE Calculate fitness of each individual in $P_{t-1}$ (Fcn.~\ref{Fcn:fitness});
		\STATE Obtain selecting probabilities (Fcn.~\ref{Fcn:prob});
		\FOR{$k = 1$ to $K$}
		\STATE Preserve the best individual in $P_{t-1}$ into $P_{t}^{(1)}$;
		\STATE Generate a random value $s\sim[0,1]$;
		\STATE Conduct selection, crossover, and mutation for generating new individuals according to $s$;
		\ENDFOR
		\ENDFOR
		\STATE Update finesses of individuals in $P_{t}$;
		\STATE Establish a new generator network $\hat{G}$ by exploiting to the best individual in $P_T$;
		\ENSURE The new generator $\hat{G}$ with the searched architecture after fine-tuning using the entire training set.
	\end{algorithmic}
\end{algorithm}
\begin{figure*}[ht]
	\vspace{0.0em}
	\centering
	\setlength{\tabcolsep}{1pt}
	\begin{tabular}{ccccc}
		\small Input Images & \small Original Results & \small $\gamma = 0.1$ & \small $\gamma = 0.01$ & \small $\gamma = 0.001$ \\
		\includegraphics[width=0.18\linewidth,height=0.09\linewidth]{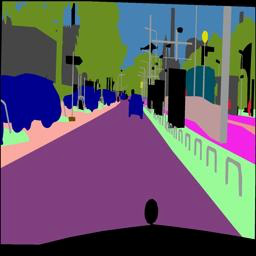} & 
		\includegraphics[width=0.18\linewidth,height=0.09\linewidth]{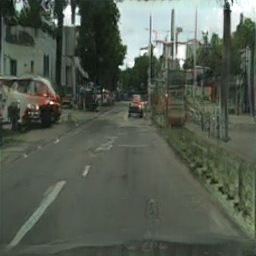} & 
		\includegraphics[width=0.18\linewidth,height=0.09\linewidth]{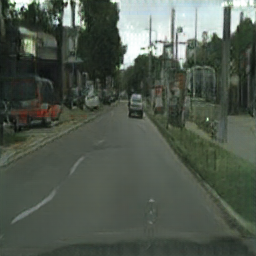} & 
		\includegraphics[width=0.18\linewidth,height=0.09\linewidth]{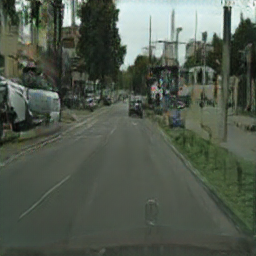} & 
		\includegraphics[width=0.18\linewidth,height=0.09\linewidth]{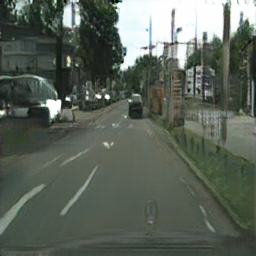} \\
		\includegraphics[width=0.18\linewidth,height=0.09\linewidth]{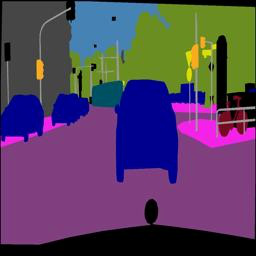} &
		\includegraphics[width=0.18\linewidth,height=0.09\linewidth]{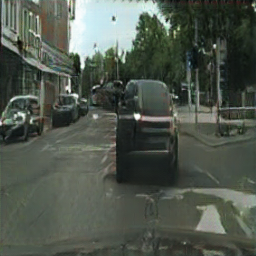} & 
		\includegraphics[width=0.18\linewidth,height=0.09\linewidth]{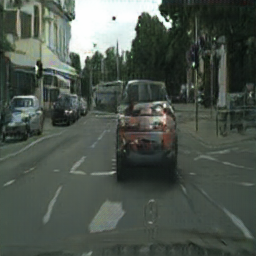} & 
		\includegraphics[width=0.18\linewidth,height=0.09\linewidth]{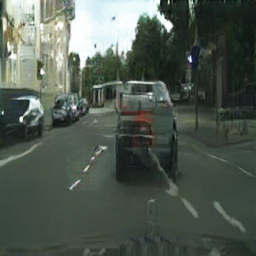} & 
		\includegraphics[width=0.18\linewidth,height=0.09\linewidth]{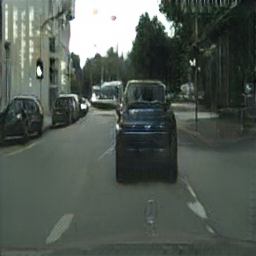} \\
	\end{tabular}	
	\vspace{0.0em}
	\caption{Images generated using the generator searched by exploiting the proposed method with different hyper-parameters $\gamma$, which balance the computation FLOPs and accumulated generator loss. A lower $\gamma$ represents model with larger FLOPs.}
	\label{Fig:params}
	\vspace{0.0em}
\end{figure*}

\section{Experiments}

In the image-to-image translation task, we evaluate the effectiveness of the proposed method to search the optimal architecture of an image generator. Extensive experiments are conducted on several paired image-to-image translator dataset including cityscapes, facades and maps. For a fair comparison, we use the same training strategy and same discriminator structure as pix2pix~\cite{pix2pix}. 

For parameters settings, we use $\lambda=100$, max generation $T=100$ and number of population is $K=32$. Probabilities of $s_1=0.2, s_2=0.7, s_3=0.1$ are adopted for selection, cross over and mutation in each generation as suggested in \cite{wang2018towards}.
For the input image of $256\times256$ and output image of $256\times256$, the length for channel number code is $L_{c_{1}}=8$, and the length for skip connection code is $L_{c_{2}}=7$. First, we conduct architecture search experiment on the cityscape dataset with different hyper parameters. Then, we immigrate the searched architecture to other image-to-image translation dataset like maps and facades which doesn’t get involved of the searching procedure to prove the generalization ability of the searched architecture.
\begin{figure*}[h]
	\centering
	\vspace{0.0em}
	\setlength{\tabcolsep}{1pt}
	\begin{tabular}{cccccc}
		\small Input Images & \small Original Results & \small Proposed Method &	\small Input Images & \small Original Results & \small Proposed method \\
		\includegraphics[width=0.15\linewidth]{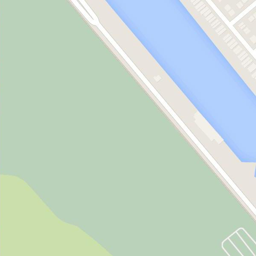} & 
		\includegraphics[width=0.15\linewidth]{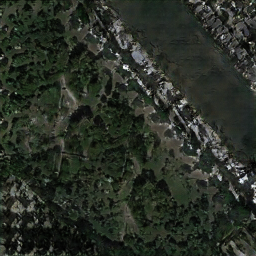} & 
		\includegraphics[width=0.15\linewidth]{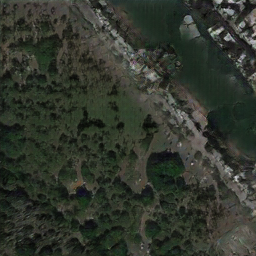} &
		\includegraphics[width=0.15\linewidth]{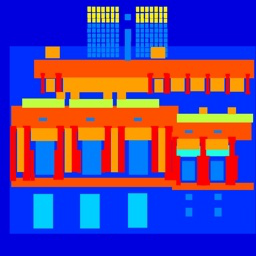} & 
		\includegraphics[width=0.15\linewidth]{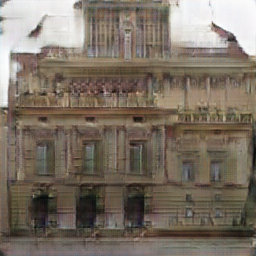} & 
		\includegraphics[width=0.15\linewidth]{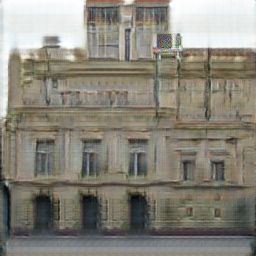} \\ 
		\includegraphics[width=0.15\linewidth]{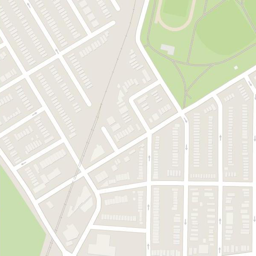} & 
		\includegraphics[width=0.15\linewidth]{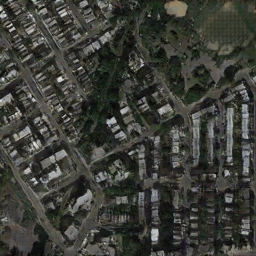} & 
		\includegraphics[width=0.15\linewidth]{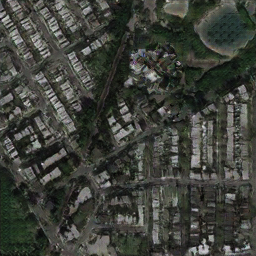} & 
		\includegraphics[width=0.15\linewidth]{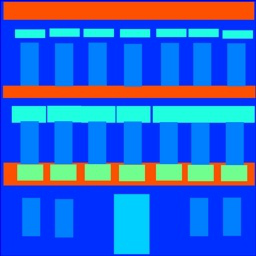} & 
		\includegraphics[width=0.15\linewidth]{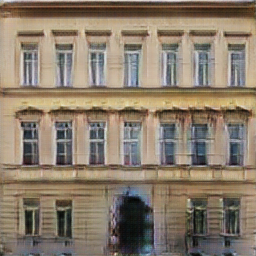} & 
		\includegraphics[width=0.15\linewidth]{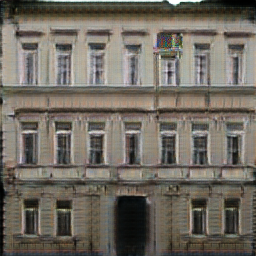} \\
	\end{tabular}
	\vspace{-0.5em}
	\caption{Some translation results on maps dataset and facades dataset respectively. The network architecture is directly borrowed from the search result of cityscapes dataset. The size of the searched model is about 22\emph{MB} compared to 208\emph{MB} of the original model, and the FLOPs of searched network is significantly fewer than that of the original one.}
	\label{Fig:maps}
	\vspace{-0.5em}
\end{figure*}

\subsection{Search on Cityscapes Dataset}
We conduct network architecture search experiments on cityscapes dataset, from semantic maps to street views. We use the described genetic algorithm to explore the defined network search space. In the experiments, we adopt different hyper parameters to balance model performance and computation cost. A smaller hyper parameter $\gamma$ means less focus on the network FLOPs, resulting in more computation consumption and better image translation quality.
\begin{table*}[htb]
	\centering
	\caption{FCN scores of different generators calculated on the cityscapes dataset. }
	\label{Tab:FCN}
	\begin{tabular}{c||c|c|c|c|c}
		\hline
		Method & Memory & FLOPs & Mean Pixel Acc & Mean Class Acc. & Mean class IoU \\
		\hline\hline
		Original~\cite{pix2pix}  &$208$\emph{MB} &18147\emph{M} &0.723 &0.244 &0.186 \\
		$\gamma=0.1$ &$54$\emph{MB}  &8422\emph{M} &0.717 &0.241 &0.184 \\
		$\gamma=0.01$  &$22$\emph{MB} &15363\emph{M} &0.738 &0.248 &0.190 \\
		$\gamma=0.001$  &$87$\emph{MB} &47431\emph{M} &0.744 &0.250 &0.197 \\
		\hline
	\end{tabular}
	\vspace{-1.0em}
\end{table*} 
\begin{figure*}[h]
	\centering
	\includegraphics[width=0.95\linewidth]{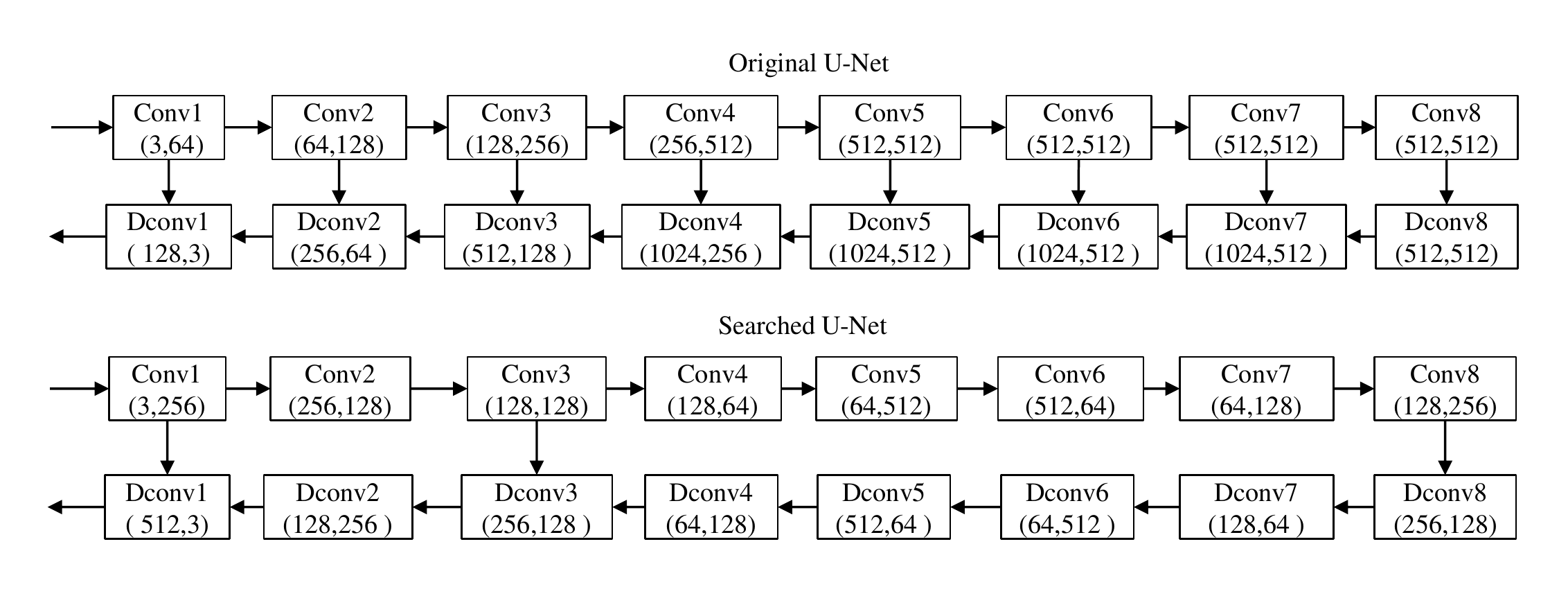}
	\vspace{-.5em}
	\caption{Comparison of the original and searched network architecture. The top shows the architecture of the original U-Net. The bottom shows the architecture of the searched U-Net with $\gamma =0.01$. The two numbers in each box represent the input channel number and output channel number of each convolutional layers on deconvolutional layers, respectively. The searched U-Net has sparse skip connections compared to the original one, and the channel numbers do not increase as the network goes deeper.}
	\label{Fig:arch}
\end{figure*} 
   
To further evaluate the effectiveness of the searching method, we conduct quantitative experiments to evaluate the performance of searched architecture. We adopt "FCN-score" proposed in \cite{pix2pix} to numerically evaluate the architecture. A pre-trained FCN-8s \cite{FCN} network is utilized to conduct segmentation task. Three measurements, mean pixel accuracy, mean class accuracy and mean class IOU are calculated. The results are shown in Tab. \ref{Tab:FCN}. The FLOPs of the model decrease with the increasing of gamma but mean pixel accuracy, mean class accuracy and mean class IOU decrease as well. When $\gamma=0.01$,  the network architecture can achieve higher FCN scores than that of the original model. The memory size of the model is only 22MB, almost 1/10 of the original model. when $\gamma=0.1$, we can get a model which costs less than half of the original model but achieves slightly lower scores. Actually, we can get a various generator architectures fit for different applications. In addition, we can conclude that the model performance is more related with computation FLOPs, rather than the memory size of the model. Interestingly, all of searched models have less parameters than that of the original model, which means that the searched architecture cost less memory storage while achieving comparable or even better performance than the original one. The parameter redundancy of original U-Net is very severe in the image-to-image translation task.
In Fig.~\ref{Fig:params}, we show some results of the image translation from semantic map to street views. With a lower $\gamma$, the number of FLOPs of model is larger, which leads to more computation cost but the better image translation quality. 

\subsection{Architecture Transfer to Other Datasets}
After network search experiments on cityscapes dataset, we get new architectures of generator. To demonstrate the effectiveness of the searched architecture, we utilize the architecture to train other paired image-to-image tasks, such as maps to satellite maps and color bar labels to real facades. For a fair comparison, we use the architecture with $\gamma=0.01$ which has less FLOPs compared to that of original U-Net to train on other image-to-image datasets, such as maps and facades.
Fig.~\ref{Fig:maps} show some results of the image transfer tasks of maps and facades respectively. From visual respect, newly searched network outperforms the original network architecture with less FLOPs and far less parameters. On maps dataset, the searched network generates satellite images more consistent with input maps. On facades dataset, the searched network generates images with more realistic windows and doors. It is worth noting that neither dataset of maps nor facades gets involved in the process of network architecture search. It demonstrates the generalization of the searched architecture and the effectiveness of the searching method furthermore.

\subsection{Discussion of the Searched Architecture}
Through the previous network architecture searching experiment, we get new architectures of generator.  As shown in Fig.~\ref{Fig:arch}, we compare the new architecture with $\gamma=0.01$ and original network architecture. These two architectures have comparable FLOPs and the new one has slightly less. In the searched architecture, we find it is not necessary to increase number of channels as network goes deeper. Only a few wide layers at specific positions is enough for good performance. Similar phenomenons can also be observed in the other two architectures in Tab.~\ref{Tab:FCN}

In addition, not every connection is essential, the first and third skip connections are preserved while other skip connections are removed. This phenomenon shows that the skip connection involving layers with larger feature maps' sizes are more essential in the image-to-image translation task.  On the other hand, it also demonstrates that long distance connection is more important than the close connection, since far skip connection mixes less correlated information whereas the close skip connection consults more computation cost while involve less important information. Thus, most of the connections are removed. 

As for the channel numbers of each convolutional layers and deconvolutional layers, the original U-Net has a similar encoder-decoder structure and gradually gets wider with continuously more convolutional layers. It reaches the widest channel numbers when feature map approaches the bottleneck. Then, the channel numbers of each subsequent deconvolutional layer is reducing accordingly. For the newly searched architecture, this principle seems to be broken. Shallow layers tend to have more channels while layers near the bottleneck tend to have less channels. More computation cost lies in the shallow layers. In the image-to-image translation task, more carefully designed shallow layers help to restore of details of the image.

\section{Conclusion}
In this paper, we propose a novel automatically network search method for paired image-to-image translator. By utilizing the evolutional algorithm, we search the channel numbers of each convolutional layers and deconvolutional layers and connections of the skip connections for a better generator architecture. We carefully design the genetic algorithm for the architecture search of U-Net. Extensive experiments with different hyper parameters are conducted to balance model performance and computation cost. Through experiments on cityscapes dataset, we get a high performance model with fewer FLOPs compared to the original backbone U-Net. The FCN scores of new architecture on cityscapes segmentation exceed the original U-Net. Furthermore, we immigrate this model architecture to other paired image-to-image translation tasks including maps and facades, and the new architecture outperforms the original one. Finally, we discuss the new architecture with the original one and this may suggest more experiences to the design of network architecture.

\clearpage
\bibliographystyle{named}
\bibliography{ijcai19}

\end{document}